\documentstyle[aps,prd,preprint,floats,epsf,epsfig]{revtex}

\begin{document}

\newcommand{\be}{\begin{equation}}
\newcommand{\ee}{\end{equation}}
\newcommand{\ba}{\begin{eqnarray}}
\newcommand{\ea}{\end{eqnarray}}

\title{Analysis of $B \to K\pi$ decays with small elastic final state 
interactions}
\author{ Claudia Isola and T. N. Pham }
\address{ Centre de Physique Théorique, \\Centre National de la Recherche 
Scientifique, UMR 7644,\\ 
Ecole Polytechnique, 91128 Palaiseau Cedex, France\\}

\date{24 August 2000}
\maketitle

\begin{abstract}
An analysis of $B \to K\pi$ decays is given, assuming a small 
elastic $\pi K $ rescattering phase difference. Using factorization model
elastic $\pi K $ rescattering phase difference. Using factorization model
only for the tree-level and electroweak penguin amplitudes, we show that
the strong penguin amplitude, its absorptive part and the CP-violation
weak phase $\gamma$ can be obtained from the measured $B \to K\pi$ 
branching ratios. The strength of the strong penguin and its absorptive
part thus obtained from the CLEO data, are found to be very close to
factorization model values and suggests a current $s$ quark mass around
$m_{s} = 106\,\rm MeV$. The central value of $\gamma$ is found to be 
around $76.10^{\circ}$, with a possible value in the range 
$50^{\circ}-100^{\circ}$.

\vspace{3cm}

\vfill

\noindent CPHT-S 085.0900  \hfill September 2000

\hfill Typeset using REV\TeX 

\end{abstract}



\narrowtext

The recent CLEO data \cite{cleo1} on charmless two-body nonleptonic
$B$ decays indicates that non-leptonic interactions enhanced by short-
distance QCD radiative corrections combined with
factorization model seems to
describe qualitatively the $B \to K\pi$ decays. In particular, the
penguin interactions
contribute a major part to the decay rates and provide an interference
between the Cabibbo-suppressed tree and penguin contribution resulting in
a CP-asymmetry between the $B \to K\pi$ decay and its charge conjugate
mode.
Though the data are not yet sufficiently accurate to allow a
determination of the weak CP-violating phase $\gamma$, they seem to
favor a large $\gamma$ in the range of $(90^{\circ}-120^{\circ})$, as
shown in a previous analysis of $B \to K\pi$ using the factorization model
with elastic rescattering phase included \cite{Isola}. This
large value of $\gamma$ is also found by the CLEO Collaboration 
in an analysis of all measured charmless two-body
$B$ decays with the factorization model \cite{cleo2}. A large $\gamma$ 
would also help to explain the suppression of the  
$\bar{B}^{0} \to \pi^{+}\pi^{-} $ decays as the interference between the
tree-level and penguin terms which increases the $B \to K\pi$ decay rates,
becomes destructive in $B \to \pi\pi$
decays. However, as shown in our previous analysis, 
a large $\gamma$ would require a large
$\pi K \to \pi K$ 
rescattering phase difference in the range $(50^{\circ}-100^{\circ})$ to   
account for the near-equality of the two largest $\bar B^0 \to K^-\pi^+$
and $B^-\to \bar K^0 \pi^-$ branching ratios. Although a large elastic
rescattering phase is not excluded by experiments, it is not expected in
high energy elastic $\pi K$ scattering which is dominated by the 
isospin-independent Pomeron
exchange amplitude so that the rescattering phase
difference $\delta = \delta_{3/2} - \delta_{1/2}$ would be small in $B$
decays. Indeed, a recent analysis of $\pi K$ elastic scattering at 
the $B$ mass \cite{Delepine} finds  
$\delta =(17 \pm 3)^{\circ}$.  
For $B \to K\pi$, from the factorization model, if  $\gamma <110^{\circ}$,
with some adjustment of form 
factors, the current $s$ quark mass and 
CKM parameters it might be possible to accommodate the  two largest 
branching ratios with a small $\delta $.
It is thus useful to study the consequence of a small $\delta $
in $B \to K\pi$ decays. Infact, as shown in Fig.1 in \cite{Isola}, for
$\delta < 50^{\circ}$, the CP-averaged $B \to K\pi$ branching ratios 
are practically
independent of $\delta $ in this range, as the $\delta$-dependent 
terms which come from the small $\cos{\delta} - 1 $
and the $\sin{\delta}$ term which is also very small. Since the strong
penguin
amplitude with its small inelastic absorptive part found in perturbative 
QCD \cite{Fleischer1,Hou} 
produce sufficiently the $B \to K\pi$ decay rates, it is likely 
that, in general, the inelastic absorptive part should not be too large
in $B \to K\pi$ decays. In this case, for  
$\delta < 50^{\circ}$, the $\delta$-dependent terms in  the 
$B \to K\pi$ decay rates would  be small  and could be therefore neglected
in the CP-averaged $B \to K\pi$ decay rates. In particular, 
for $\delta$ in this range, the 
$B^{-} \to \bar{K}^{0}\pi^{-}$ decay rate,  is 
essentially given by the $\gamma$-independent strong penguin
contribution. Now, if  we assume factorization for the small 
tree-level and electroweak terms, the strong penguin and its
absorptive part and $\gamma$ could then be obtained from the measured
CP-averaged $B \to K\pi$ decay rates. With the dominant
strong penguin contribution obtained from experiments, 
the value of $\gamma$ thus obtained  
is subjected to very little theoretical uncertainties in contrast with 
the result from the factorization model for the whole
amplitude as the penguin matrix elements are sensitive to the 
$s$ quark mass which is not known to a good accuracy. This is the main
purpose of this paper. In the following, 
we shall obtain the strong
penguin contribution and $\gamma$ from the measured $B \to K\pi$
decay rates, assuming factorization for the tree-level and electrowaek
penguin terms and a small $\pi K$ rescattering phase difference $\delta$.
We find agreement with factorization model for the strong penguin 
contribution and its absorptive part. The value of $\gamma$ is found 
to be in the range $(50^{\circ}-100^{\circ})$, with a central 
value of $70^{\circ}$. To proceed, we begin by writing
down the $B \to K\pi$ decay amplitudes in terms of the decay amplitudes
into $I=1/2$ and $I=3/2$ final states \cite{Isola,Deshpande},
\begin{eqnarray}
A_{K^-\pi^0} &=& 
 {2\over 3} B_3e^{i\delta_3} + \sqrt{{1\over 3}} 
(A_1+B_1)e^{i\delta_1}, \ \ \  \nonumber\\
A_{\bar K^0\pi^-} &=&
{\sqrt{2}\over 3} B_3e^{i\delta_3} - \sqrt{{2\over 3}} 
(A_1+B_1)e^{i\delta_1},\nonumber\\
A_{K^-\pi^+} &=&
{\sqrt{2}\over 3} B_3e^{i\delta_3} + \sqrt{{2\over 3}} 
(A_1-B_1)e^{i\delta_1},\ \ \ \nonumber\\
A_{\bar K^0\pi^0} &=& 
{2\over 3} B_3e^{i\delta_3} - \sqrt{{1\over 3}} 
(A_1-B_1)e^{i\delta_1},
\label{amplitude}
\end{eqnarray}
$A_{1}$ is the sum of  the strong penguin $A_{1}^{\rm S} $
and the $I=0$  tree-level $A_{1}^{\rm T}$ as well as the $I=0$ 
electroweak penguin $A_{1}^{\rm W}$ contributions 
to the $B \to K\pi$ $I=1/2$ amplitude;  similarly  
$ B_{1}$ is the sum of the
$I=1$ tree-level $B^{T}_{1}$ and electroweak penguin $B^{W}_{1}$
contribution to the $I=1/2$ amplitude, and $B_{3}$ is the sum of the
$I=1$  tree-level $B^{T}_{3}$ and electroweak penguin $B^{W}_{3}$
contribution to the $I=3/2$ amplitude. Using the following 
effective Hamiltonian for $B \to K\pi$ decays 
\cite{Fleischer1,Buras,Ciuchini,Kramer,Deshpande1} 
\begin{eqnarray}
H_{\rm eff} &=& {G_F\over \sqrt{2}}[V_{ub}V^*_{us}(c_1O_{1}^{u} + c_2
O_{2}^{u}) + V_{cb}V^*_{cs}(c_1O_{1}^{c} + c_2 O_{2}^{c})  \nonumber\\
&&-\sum_{i=3}^{10}(V_{tb}V^*_{ts} c_i) O_i] 
+ {\rm h.c.}\,,
\label{hw}
\end{eqnarray}
where the $c_{i}$,  at the next-to-leading logarithms, take the form 
of an effective
Wilson coefficients $c_{i}^{\rm eff}$ which  contain  also the penguin 
contribution from the $c$ quark loop \cite{Fleischer1,Deshpande1},
we obtain in the factorization model,
\ba
&&A_1^{\rm T} = i{\sqrt{3}\over 4}\, V_{ub}V_{us}^*\, r\, 
a_2,\nonumber\\
&&B_1^{\rm T} = i {1\over 2\sqrt{3}}\, V_{ub}V_{us}^*\, r 
\left[-{1\over 2}a_2 
+a_1 X\right],\nonumber\\
&&B_3^{\rm T} = i{1\over 2}\, V_{ub}V_{us}^*\, r 
\left[a_2 + a_1 X\right],
\nonumber\\
&&A_1^{\rm S} = -i{\sqrt{3}\over 2}\, V_{tb}V_{ts}^*\, r
\left[ a_4 + a_6 Y\right],
\ \ \ \ \ \ B_1^{\rm S} = B_3^{\rm S} = 0\nonumber\\
&&A_1^{\rm W} = -i{\sqrt{3}\over 8}\, V_{tb}V_{ts}^*\, r
\left[a_8Y +  a_{10}\right],
\nonumber\\
&&B_1^{\rm W} = i{\sqrt{3}\over 4}\, V_{tb}V_{ts}^*\, r
\left[{1\over 2}a_8Y 
+ {1\over 2}a_{10} 
+\left(a_7 - a_9\right)X\right],
\nonumber\\
&&B_3^{\rm W} = - i{3\over 4}\, V_{tb}V_{ts}^*\, r
\left[\left( a_8Y 
+ a_{10} \right) 
-\left(a_7 - a_9\right)X\right],
\label{AB1}
\ea
where $r = G_F\, f_K F^{B\pi}_0(m^2_K) (m_B^2-m_\pi^2)$,
$X= (f_\pi/f_K)(F^{BK}_0(m^2_\pi)/F^{B\pi}_0(m^2_K)) 
(m_B^2-m^2_K)/(m_B^2-m_\pi^2)$, 
$Y = 2m^2_K/[(m_s+ m_q)(m_b-m_q)]$ with $q=u,\ d$ for 
$\pi^{\pm,0}$ final states, respectively. In this analysis, 
$f_{\pi} =133 \,\rm MeV$, $f_{K} =158\,\rm MeV$, 
$ F_{0}^{B\pi}(0) = 0.33$, $ F_{0}^{BK}(0) = 0.38$ \cite{Ali,Bauer}~;
$|V_{cb}|=0.0395, |V_{cd}|=0.224 $ and $|V_{ub}|/|V_{cb}|=0.08$
\cite{PDG}.
We take $m_{s} =120\,\rm MeV$. The value of $m_{s}$ is not known to a 
good accuracy, but a value around $(100-120\rm )\, MeV$
inferred from $m_{K^{*}} - m_{\rho}$, $m_{D_{s}^{+}}- m_{D^{+}}$ 
and $m_{B_{s}^{0}}- m_{B^{0}}$ mass differences \cite{Colangelo} 
seems not unreasonable. $a_{j}$ are given in terms of the effective 
Wilson coefficients
$c^{\rm eff}_{j}$ ($N_{c}$ is the number of effective colors) by
\ba
a_{j} &=& c^{\rm eff}_{j} + c^{\rm eff}_{j+1}/N_{c} \,\, {\rm for} \
j=1,3,5,7,9 \nonumber\\
a_{j} &=& c^{\rm eff}_{j} + c^{\rm eff}_{j-1}/N_{c} \,\, {\rm for} \
j=2,4,6,8,10\,\, .
\ea
For $N_c =3$ and $m_{b} = 5.0\, \rm GeV$ ,
$a_{j}$  take the following numerical values
\ba
a_1 &=& \ 0.07    , \ \ \ a_2=  \ 1.05  , \nonumber\\
a_4 &=& -0.043 - 0.016i ,\ \ \ a_6= -0.054 - 0.016i  , \nonumber\\
a_7 &=& \ 0.00004 -0.00009i  , \ a_8 =  0.00033 - 0.00003i  , \nonumber\\
a_9 &=& \-0.00907-0.00009i  , \ a_{10}=  -0.0013 -0.00003i .
\label{coeff}
\ea
As pointed out in Ref.\cite{Isola}, $a_{1}$ is sensitive to  
$N_{c}$ and is rather  small for 
$N_{c} =3$ . As there is no evidence for a large positive $a_{1}$ in
$B \to K\pi$ decays which are penguin-dominated and are not sensitive 
to $a_{1}$, we use $a_{1}$ evaluated with $N_{c} =3$ given in 
Eq.(\ref{coeff}). Indeed, the predicted branching ratios remain
essentially unchanged with $a_{1}=0.20$ taken from the Cabibbo-favored 
$B$ decays \cite{Kamal1}.  
 We note that the coefficients 
$c^{\rm eff}_{3}$,
$c^{\rm eff}_{4}$, $c^{\rm eff}_{5}$ and $c^{\rm eff}_{6}$, 
are enhanced by the internal charm quark loop  due to the large
time-like virtual gluon momentum $q^{2} = m_{b}^{2}/2$ as pointed out
in \cite{Fleischer1,Hou} (the other electroweak penguin 
coefficients like $c^{\rm eff}_{7}$  and $c^{\rm eff}_{9}$ are not
affected by the charm quark loop contribution in any significant amount).
This enhancement of the strong penguin term increases the decay rates and 
bring the theoretical $B \to K\pi$ decay rates closer to the latest
CLEO measurements.  
The  tree-level amplitudes are suppressed
relative to the penguin terms by the CKM factor
$V_{ub}V_{us}^*/V_{tb}V_{ts}^* $ which can be approximated by
$-(|V_{ub}|/|V_{cb}|)\times (|V_{cd}|/|V_{ud}|)\exp(-i\,\gamma)$
after neglecting terms of the order $O(\lambda^{5})$ in the (bs)
unitarity triangle.   
The $B \to K\pi$ decay rates then depend on the FSI rescattering
phase difference $\delta = \delta_{3} -\delta_{1}$ and the weak phase
$\gamma$. 
The factorization model prediction for $B \to K\pi$ decys is shown in 
Fig.1 and Fig.4 of \cite{Isola}, where  the $B \to K\pi$ branching 
ratios are plotted against $\delta$ for $\gamma =70^{\circ}$ and  
$\gamma =110^{\circ}$ respectively. As can be seen, factorization 
with the short-distance Wilson coefficients obtained by perturbative
QCD reproduces qualitatively the CLEO data, which gives \cite{cleo1}
\begin{eqnarray}  
{\mathcal B} (B^{+} \rightarrow K^{+} \pi^0)
&=&(11.6^{+3.0+1.4}_{-2.7-1.3}) \times 10^{-6},  \nonumber  \\
{\mathcal B} (B^{+} \rightarrow K^0 \pi^{+})  &=& 
(18.2^{+4.6}_{-4.0}\pm 1.6) \times 10^{-6},   \nonumber  \\
{\mathcal B} (B^0\rightarrow K^{+} \pi^{-})  &=& 
(17.2^{+2.5}_{-2.4}\pm 1.2) \times 10^{-6}, \nonumber \\ 
{\mathcal B} (B^0 \rightarrow K^0 \pi^0)  &=&
(14.6^{+5.9+2.4}_{-5.1-3.3}) \times 10^{-6}.
\end{eqnarray}
 We remark that for 
$\delta < 50^{\circ}$, the $B \to K\pi$ branching ratios show
practically no variation with $\delta$ which, as explained earlier, come
from the small $\cos(\delta) - 1 $ and the $\sin(\delta)$ term. Also 
the computed branching ratios for 
$\gamma = 110^{\circ}$ are somewhat larger than 
the values with $\gamma = 70^{\circ}$. Factorisation also shows that  
$B^-\to \bar K^0 \pi^-$
and $\bar B^0 \to K^-\pi^+$ are the two largest modes, in qualitative 
agreement with the CLEO data which give  near-equality of these  two
largest branching ratios.    
Fig.1 of \cite{Isola} shows that the two largest branching ratios 
are quite apart,
except for $\delta < 50^{\circ}$ where the difference of these two
branching ratios becomes smaller, about  $2.0\times 10^{-6}$. For   
larger $\gamma $, as in Fig.4 of \cite{Isola}, this difference 
reverses the sign and 
become large for $\gamma = 110^{\circ}$, even for $\delta < 50^{\circ}$,
except in a small region around $\delta = 50^{\circ}$. As mentioned
earlier, this value of $\delta $ seems too large compared with a 
theoretical value of $(17 \pm 3)^{\circ}$ \cite{Delepine}. We thus have
to accomodate the $B \to K\pi$ data with  $\delta < 50^{\circ}$
and  $\gamma < 110^{\circ}$. We now assume that $\delta < 50^{\circ}$
and proceed to the determination of the strong penguin contribution and
the weak phase $\gamma$ from the CP-averaged $B \to K\pi$ branching
ratios assuming factorization for the tree-level and electroweak penguin
contributions. A test of factorization for the strong penguin
contribution could then be made by comparing the value obtained from
experiments with the factorization prediction. The value for $\gamma $
thus obtained will not suffer from the uncertainties in the 
computation of the strong penguin matrix elements. For this purpose we 
need the $B^-\to \bar K^0 \pi^-$ branching ratio for which the 
$\delta $-dependent terms are neglected and two other $\delta
$-independent quantitiies obtained from the $B^-\to \bar K^0 \pi^-$, 
$B^-\to K^{-} {\pi}^{0} $ and $\bar B^0\to K^- {\pi}^+ $ decay rates
given by \cite{Isola}
\ba
Q_{12} &=& r_b\left[{\mathcal B}(B^-\to K^{-} {\pi}^{0})
+{\mathcal B}( B^-\to \bar{K}^{0} {\pi}^{-})\right]\nonumber\\
Q_{23} &=& r_b{\mathcal B}(B^-\to\bar{K}^{0}
    {\pi}^{-})+{{\mathcal B}(\bar B^0\to K^-
      {\pi}^+)}\nonumber\\
Q_{2}  &=&r_b\,{\mathcal B}( B^-\to \bar{K}^{0} {\pi}^{-})_{\delta=0}
\label{input1}
\ea 
where $r_b=\tau_{B^0}/ \tau_{B^-}$.
Similarly,
\ba
Q_{34} &=& {\mathcal B}(\bar{B}^{0} \to K^- {\pi}^+) + {\mathcal B}
(\bar{B}^{0} \to \bar{K}^{0} {\pi}^{0})\nonumber\\
Q_{14} &=& r_b{\mathcal B}(B^-\to K^- {\pi}^0)
    +{\mathcal B}(\bar B^0\to \bar{K}^{0} {\pi}^{0})\,\,.
\label{input2}
\ea 
 
The strong penguin amplitude $A_1^{\rm S}$, in the factorization model,
is proportional to $ a_4 + a_6 Y $. We now consider 
the quantity $ a_4 + a_6 Y $ as a parameter
and write $P\exp(i\,h) = -10\sqrt{3}\,(a_4 + a_6 Y)/2$, $P\sin{h}$ is then
the 
absorptive part of the strong penguin amplitude. $A_1^{\rm S}$ is now
given by
\be
A_1^{\rm S} = i{1\over 10}\, V_{tb}V_{ts}^*\, r\,P\exp(i\,h),
\label{AS1}
\ee 
where we have parametrised the strong penguin amplitude in terms of 
the quantity $r$ as in the factorization model. With factorization for
the tree-level and electroweak penguin amplitude, the $B \to K\pi$ decays
branching ratios can now be obtained in terms of 3 parameters, $P$, 
the inelastic strong phase $h$ and $\gamma$ which can now be determined
from experiments. In Eq.(\ref{input1}) , $Q_{2}$  is the 
$ B^-\to \bar{K}^{0} {\pi}^{-}$ branching ratios(CP-averaged) evaluated
for $\delta =0$, at which the tree-level term vanished. $Q_{2}$ 
gives us the strength of the penguin contribution $P$. $Q_{12}$ and  
$Q_{23}$,  
can then determine $P\cos{h}$ and $\cos{\gamma}$. As the experimental
errors
is larger in ${\mathcal B}(\bar{B}^{0} \to \bar{K}^{0}\pi^{0})$, we have
used $Q_{23}$ which is also  
$\delta $-indepedent to a good approximation, instead of the 
quantity $Q_{34}$ which is  the sum of 
${\mathcal B}(\bar B^0\to K^- {\pi}^+) $ and 
${\mathcal B}(\bar{B}^{0} \to \bar{K}^{0}\pi^{0})$. With the numerical
values for the effective Wilson coefficients of the tree-level and
electroweak operators and the BWS value for the form factors
\cite{Bauer}, we find, for the CP-averaged branching ratios,
\ba
Q_{12} &=& (3.984\,P^{2} + 0.301\,P\cos{h} - 0.459\,P\cos{h}\cos{\gamma}
\nonumber\\
&&-0.057\,\cos{\gamma} + 0.060)\times 10^{-5}
\nonumber\\
Q_{23} &=& (5.312\,P^{2} + 0.030\,P\cos{h} -
0.862\,P\cos{h}\cos{\gamma}\nonumber\\
&&-0.010\,\cos{\gamma} + 0.070)\times 10^{-5}
\nonumber\\
Q_{2}  &=&(2.656\,P^{2} - 0.030\,P\cos{h})\times 10^{-5}\,\,. 
\label{n1}
\ea 
Similarly, we obtain,
\ba
Q_{34} &=& (3.984\,P^{2}  - 0.256\,P\cos{h} - 0.834\,P\cos{h}\cos{\gamma}
\nonumber\\
&&- 0.013\,\cos{\gamma} + 0.089)\times 10^{-5} 
\nonumber\\
Q_{14} &=& (2.656\,P^{2} + 0.015\,P\cos{h} - 0.431\,P\cos{h}\cos{\gamma}
\nonumber\\
&&-0.060\,\cos{\gamma} + 0.079)\times 10^{-5}\,\,.
\nonumber\\
\label{n2}
\ea 
In the expressions for $Q_{23}$ and $Q_{14}$ shown above,
a negligible $\cos \delta \cos \gamma$ terms of the order $10^{-7}$
has been included for completeness. 
Comparing $Q_{23}$ and $Q_{14}$ in Eq.(\ref{n1}) and Eq.(\ref{n2}),
we find,   
\be
{ r_b}{\mathcal B}_{\bar{K}^{0}
    {\pi}^{-}}+{\mathcal B}_{K^- {\pi}^+} = 
2\,\left[{\mathcal B}_{\bar{K}^{0} {\pi}^{0}}+
{ r_b}{\mathcal B}_{ K^- {\pi}^0}\right]
\label{rel}
\ee
valid up to a  small $\delta$-dependent term $\Delta$ ($C$ being the 
usual phase space factor)
\ba
\Delta & = &\left\{ \Gamma(B^{-} \to \bar{K}^{0}
{\pi}^{-})+\Gamma(\bar{B}^{0}
\to K^- {\pi}^+) \right. \nonumber\\
   & - & \left. 2\left[\Gamma(B^{-} \to K^- {\pi}^0)+\Gamma(\bar{B}^{0}
\to
     \bar{K}^{0} {\pi}^{0})\right]\right\}\tau_{B^0}\nonumber\\
& = & \left[-{4\over 3}|B_3|^2-{8\over {\sqrt 3}}{\rm Re} (B_3^* B_1 {\rm
  e}^{i\delta})\right](C\tau_{B^0})\,\, 
\label{delta}
\ea
which is of the order  $O(10^{-6})$. Eq.(\ref{rel}) then gives, to a
good approximation
\be
{\mathcal B}_{\bar{K}^{0}{\pi}^{0}}=(1/2)(r_b{\mathcal
B}_{\bar{K}^{0} {\pi}^{-}}+{\mathcal B}_{K^- {\pi}^+}) - { r_b}{\mathcal
B}_{ K^- {\pi}^0}\,\,.
\label{kpi0}
\ee
The CLEO data \cite{cleo1} then gives, 
\be
{\mathcal B}_{\bar{K}^{0}{\pi}^{0}}= (0.60^{+0.7}_{-0.6})\times
10^{-5}\,\,.
\label{kpi01}
\ee
The above predicted central value is 
smaller than the CLEO value, but a more
precise test of this relation must await further measuremnts of the
$B \to K\pi$ decays, when a more accurate value for the 
$\bar{B}^{0} \to \bar{K}^{0}\pi^{0}$ branching ratio. For this reason,
we shall not use the measured value for    
${\mathcal B}_{\bar{K}^{0}{\pi}^{0}} $ and use only the 3 quantities 
$Q_{2}$, $Q_{12}$ and  $Q_{23}$ given in Eq.(\ref{n1}) in our
determination of $P$, $h$ and $\gamma$  in this analysis. We note that 
$Q_{2}$, which is almost independent of $\gamma$ and $P\cos h$, allows a
determination of the strength of the strong penguin intereactions $P$. 
Infact, by neglecting terms of the order $10^{-6}$ or smaller in
$Q^{2}$, we find, using the measured value for 
${\mathcal B} (B^{+} \rightarrow K^0 \pi^{+})$, 
\be
P= 0.81\pm 0.12
\label{P}
\ee
to be compared with 
\be
P_{0}= 0.77
\label{P0}
\ee
obtained from
perturbative QCD and factorization model with $m_{s} = 120\,\rm MeV$
and normalised with the BSW form factor \cite{Isola}. This 
suggests that factorization model could accommodate the 
$B^{+} \rightarrow K^0 \pi^{+} $ branching ratio with 
$m_{s} =106\,\rm MeV$. It is more difficult to obtain $P\cos h$ 
and $P\cos h\cos \gamma$ as these terms are present in the decay rates
with small coefficients and the large experimental errors in the current
measured branching ratios. However,  if we consider the quantity
$D_{23}= Q_{23}- 2Q_{2}$ and $D_{12}= Q_{12}- 1.5\,Q_{2}$ obtained
from Eq.(\ref{n1}), as given by,
\ba
D_{23} &=&( -0.86\,P\cos h\cos \gamma + 0.09\,P\cos h + 0.07)
\times 10^{-5}\nonumber\\
D_{12} &=& (-0.459\,P\cos h\cos \gamma + 0.345\,P\cos h \nonumber\\
&&+0.057\,\cos{\gamma} + 0.060)\times 10^{-5}
\label{d23}
\ea
we can infer from the data which show  the near-equality of 
the two largest branching ratios $ {\mathcal B}_{\bar{K}^{0}{\pi}^{-}}$
and ${\mathcal B}_{K^- {\pi}^+} $  that
$P\cos h$ is $O(1)$
and $\cos \gamma$ is smaller, of the order $O(10^{-1})$. This shows that 
the absorptive part of the strong penguin contribution is not large and  
the weak phase angle $\gamma$ should be around $90^{\circ}$.
Indeed, by solving Eq.(\ref{n1}) for $P^{2} $, $P\cos h $ 
and $P\cos h \cos \gamma$
with the central values for the measured quantities $Q_{2}$, $Q_{12}$
and $Q_{23}$, we find,
\be
P^{2} = 0.668, \,P\cos h = 0.826, \, \cos \gamma = 0.240
\label{sols}
\ee
which gives,
\be
P = 0.817, \quad \cos h = 1.01, \quad  \gamma = 76.10^{\circ}\,\,.
\label{solsb}
\ee
The value for $P$, as mentioned, is quite closer to 
value of $0.77 $
obtained from perturbative QCD and factorization model with
$m_{s} = 120\,\rm MeV$. It is not surprising to obtain a value for
$\cos h$ slightly unphysical, because of large errors in the measured
branching ratios. This value indicates that $\sin h$ should be small
and hence a small absorptive part for the strong penguin contribution. 
This is consistent with the theoretical value obtained with 
perturbative QCD and factorization which gives $\cos h=0.95$.  
As the current measured values for the branching ratios have large
experimental uncertainties, of the order few times $10^{-6}$, it is not
very meaningful to quote the experimental errors in the determination of
$\gamma$. 
The value of $P$ is better determined, with an errors of about $10\%$.
Taking account of large experimental errors, we can say that $h$ should be
around $17^{\circ}$ as suggested by QCD and 
$\gamma$ should be in the range $(50-100)^{\circ}$. Infact, as shown 
in Fig.1, where $ D_{12}$ and $ D_{23}$ obtained with $\cos h =0.95$ 
is plotted against $\gamma$, the very small CLEO measured
values for  $ D_{23}$ and $ D_{12}$ would suggest a possible value of
$\gamma$ in the range $(50-100)^{\circ}$.

Taking the central value $P=0.817 $ and $\gamma = 76.10^{\circ}$ 
obtained above and $\cos h=0.95$ as suggested by QCD, we
give in Fig.2 the CP-averaged $B \to K\pi$ branching ratios plotted 
against the rescattering phase difference $\delta$. As can be seen, for 
$\delta$ below $50^{\circ}$, as assumed in this analysis, the CP-averaged
branching ratios show no visible variation with $\delta$, consistent
with our assumption that we can put $\delta =0$ in the computation of
the CP-averaged branching ratios.

The CP-asymmetries in $B \to K\pi$
decays can now be obtained by including the  $\delta$-dependent
terms in the branching ratios.  As shown in Fig.3, we find that the
asymmetries can be
appreciable, in the range $10-15\%$ for the absolute CP-asymmetries,
except
for the $\bar B^0 \to \bar K^0\pi^0$ mode, for which the asymmetries
could amount to $20-25\%$. 

In conclusion, we have shown that, as long as the elastic $\pi K$
rescattering phase difference is less than $50^{\circ}$, a determination
of the strong penguin contribution as well as its absorptive part
could be done with the measured CP-averaged $B \to K\pi$ branching ratios
using only factorization for the tree-level and electroweak penguin
amplitudes as theoretical input. We
have found that the strength of the strong penguin amplitude and its
absorptive part are very close to perturbative QCD and factorization and
suggests a value of $m_{s} = 106\,\rm MeV$. The value for $\gamma$
is found to be in the range $50^{\circ}-100^{\circ}$. With small
rescattering phase difference, we found that the CP-asymmetries are
in the range $10-15\%$.
More precise data  is needed for a precise determination of $\gamma$
without the use of factorization model for the strong penguin matrix 
elements, as done in this analysis.

\newpage
\begin{figure}[hbp]
\centering
\leavevmode
\epsfxsize=6.5cm
\epsffile{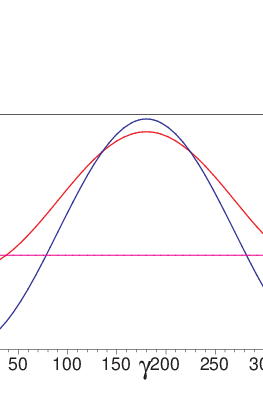}
\caption{ The curves (a) for $D_{23}$ and (b) for $ D_{12}$ versus
$\gamma$}
\label{fig:d23}
\end{figure}

\begin{figure}[hbp]
\centering
\leavevmode
\epsfxsize=6.5cm
\epsffile{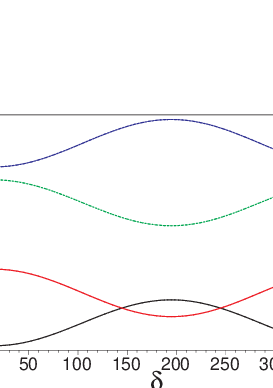}
\caption{${\mathcal B}(B\to K\pi)$ vs. $\delta$ for $\gamma = 76^\circ$.
The curves  
(a), (b), (c), (d)  are for the CP-averaged branching ratios
$B^-\to K^-\pi^0,\ \bar K^0 \pi^-$ and 
$\bar B^0 \to K^-\pi^+,\ \bar K^0\pi^0$, respectively.}
\label{fig:BRc}
\end{figure}

\begin{figure}[hbp]
\centering
\leavevmode
\epsfxsize=6.5cm
\epsffile{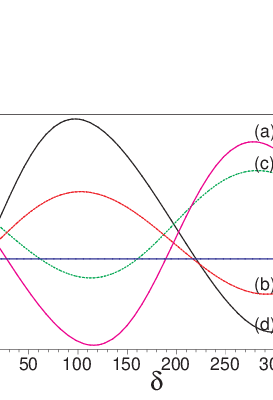}
\caption{The asymmetries vs.$\delta$ for $\gamma = 76^\circ$. 
The curves 
(a), (b), (c), (d)  are for $As_{B^-\to K^-\pi^0}$,$As_{B^-\to\bar
 K^0 \pi^-}$, $As_{\bar B^0 \to K^-\pi^+}$, $As_{\bar B^0 \to \bar
K^0\pi^0}$,}
\label{fig:asymm}
\end{figure}

\end{document}